\documentclass[prl,reprint,superscriptaddress,showpacs,showkeys,preprintnumbers,nofootinbib]{revtex4-1}
\pdfoutput=1   
\usepackage{amsmath,amsfonts,amssymb,amsbsy}
\usepackage{mathrsfs,latexsym}
\usepackage{graphicx}
\usepackage{color}
\usepackage{hyperref}
\usepackage{macros_alpha}
\usepackage{dcolumn}

\begin{document}

\preprint{WUP 14-11}
\preprint{DESY 14-190}
\preprint{SFB/CPP-14-78}

\title{%
On the effects of heavy sea quarks at low energies
}
\date{\today}
\collaboration{ALPHA collaboration}

\author{Mattia~Bruno}
\affiliation{John von Neumann Institute for Computing (NIC), DESY, Platanenallee~6, 15738 Zeuthen, Germany}
\author{Jacob~Finkenrath}
\affiliation{Department of Physics, Bergische Universit{\"a}t Wuppertal, Gaussstr.~20, 42119~Wuppertal, Germany}
\author{Francesco~Knechtli}
\affiliation{Department of Physics, Bergische Universit{\"a}t Wuppertal, Gaussstr.~20, 42119~Wuppertal, Germany}
\author{Bj{\"o}rn~Leder}
\affiliation{Department of Physics, Bergische Universit{\"a}t Wuppertal, Gaussstr.~20, 42119~Wuppertal, Germany}
\affiliation{Institut~f\"ur~Physik, Humboldt-Universit\"at~zu~Berlin, Newtonstr.~15, 12489~Berlin, Germany}
\author{Rainer~Sommer}
\affiliation{John von Neumann Institute for Computing (NIC), DESY, Platanenallee~6, 15738 Zeuthen, Germany}
\affiliation{Institut~f\"ur~Physik, Humboldt-Universit\"at~zu~Berlin, Newtonstr.~15, 12489~Berlin, Germany}

\begin{abstract}
We present a factorisation formula for the 
dependence of light hadron masses and low energy hadronic scales on the mass 
$M$ of a
heavy quark: apart from an overall factor $Q$, ratios such 
as $r_0(M)/r_0(0)$ are computable in perturbation theory at large $M$.
The mass-independent factor $Q$ is obtained from the theory in the limit $M\to0$ and
the decoupled theory with the heavy quark removed. 
The perturbation theory part is stable concerning different loop orders
and our non-perturbative
results match on quantitatively to the perturbative prediction. 

Upon taking ratios of different hadronic scales at the same mass, 
the perturbative function drops out 
and the ratios are given by the decoupled theory up to $M^{-2}$
corrections. Our present numerical results are obtained in a model calculation
where there are no light quarks and a heavy doublet of quarks is
decoupled. They are limited to masses a factor two below the charm. This is not
large enough to see the $M^{-2}$ scaling predicted by the
theory, but it is sufficient
to verify -- in the continuum limit -- that the 
sea quark effects of quarks with masses 
around the charm mass are very small.
\end{abstract}


\maketitle

\newcommand{\Lameff}{\Lambda_\mathrm{dec}}
\newcommand{\gbareff}{\overline{g}_\mathrm{dec}}
\newcommand{\gbarnl}{\overline{g}_{\tl}}
\newcommand{\gbarnf}{\overline{g}_{\tq}}
\renewcommand{\bzero}{b_0(\nq)}
\newcommand{\bzerop}{b_0(\nl)}
\newcommand{\bone}{b_1(\nq)}
\newcommand{\bonep}{b_1(\nl)}

\newcommand{\nq}{N_\mathrm{q}}
\newcommand{\tq}{\mathrm{q}}
\newcommand{\tl}{\mathrm{l}}
\newcommand{\Lamq}{\Lambda_{\tq}}
\newcommand{\Laml}{\Lambda_{\tl}}
\newcommand{\logML}{\tau}

\section{Introduction}

One usually presumes that the low energy dynamics of QCD, 
such as the hadron mass spectrum, is rather insensitive to 
the physics of heavy quarks. One can then work with 
QCD with just the three or four light quarks in order to understand
it.\footnote{Of course, in higher energy processes the heavier quarks
play a relevant role, e.g.\ in the fundamental parameters of QCD for
LHC physics, or more generally the $\Lambda$-parameter of the 5-flavor
theory.}
While large $N_\mathrm{c}$ (color) arguments suggest a general 
suppression of quark loop effects, and then a particular one
for heavy quarks, so far there has not been any 
non-perturbative investigation determining the typical magnitude
of these effects. This is understandable, since 
lattice gauge theory with heavy quarks generically has
enhanced discretisation errors and it is a non-trivial
task to separate the physical effects from those unwanted 
errors.  
It is thus of high interest for the lattice community to understand whether 
it is already time to include a charm sea quark in the simulations.
Note that one has to be precise about the meaning of the
decoupling of heavy quarks \cite{Appelquist:1974tg,Weinberg:1980wa}. 
They do leave traces through renormalisation, which we discuss below. 

The theoretical tool to understand these questions is the low
energy effective theory \cite{Weinberg:1978kz,Weinberg:1980wa} 
describing the physics with one or more 
heavy quarks decoupled. We denote this theory by \qcdm.
The leading order effective theory 
is just QCD with one or more quark flavors less. The gauge coupling $\gbareff$
 and quark masses of decQCD are adjusted such that \qcdm\ 
(approximately) reproduces 
the physics of the (more) fundamental theory at an energy sufficiently below the mass of the decoupled quark
\cite{thresh:BeWe}. This adjustment is referred to as matching.

We consider the situation with $\nl$ light quarks and 
$\nq$ quarks in total. Indicating the flavor content $\nf$ of the theory by a subscript, 
the fundamental theory is ${\rm QCD}_{\nq}$. The theory with only the light 
quarks is ${\rm QCD}_{\nl}$. Hadronic quantities, the couplings and the $\Lambda$-parameters
in these theories are distinguished by subscripts $\tq$ and $\tl$.

In this letter we briefly present the effective theory from the non-perturbative
point of view, discuss the perturbative matching of its parameters in terms of 
renormalisation group invariants (RGI) and point out the factorisation 
formula 
\bes
  \label{e:theequ}
  {\mhad_\tq(M) \over \mhad_\tq(0)} =  
    Q^\mathrm{had}_{\tl,\tq} \times {P_{\tl,\tq}(M/\Lamq)} + \rmO((\Lamq/M)^2) \,.
\ees
It gives the mass-dependence of hadron masses or hadronic scales
such as $r_0$ \cite{pot:r0} or $t_0,w_0$ \cite{flow:ML,flow:w0} in terms 
of two factors. The first factor, $Q^\mathrm{had}_{\tl,\tq}$, depends on the 
hadron mass or hadronic scale and involves only information
from the theories with $\nq$ and $\nl$  mass-less quark flavors.%
\footnote{
We here use the language of a theoretical situation where all light quarks
are mass-less. 
Light quark masses can be added with trivial changes, such as 
additional arguments in several functions.} The second factor, 
$P_{\tl,\tq}(M/\Lamq)$, gives the relation
of the  $\Lambda$-parameters of these two theories, determined 
such that the low energy physics of the fundamental theory, ${\rm QCD}_{\nq}$ with
$\nq-\nl$ quarks of RGI mass $M$, is the same as the one of ${\rm QCD}_{\nl}$
up to power corrections $\rmO((\Lamq/M)^2)$.
Throughout this letter we take the $\Lambda$-parameters to be defined in the 
$\msbar$ scheme, but this choice is irrelevant, namely $Q,P$ have a trivial 
scheme dependence in regular schemes.\footnote{In regular schemes
the couplings are related to, say, the $\msbar$ one by 
$\gbar^2 = \gbar^2_\msbar +\rmO(\gbar^4_\msbar)$.} Interestingly,
the asymptotics of $P_{\tl,\tq}(M/\Lamq)$ for large mass $M$, 
is computable in perturbation theory. The formula thus provides 
a factorisation into a non-perturbative piece, $Q$, and 
a ``perturbative'' one. In particular, the  mass-dependence is ``perturbative''.
We here use quotation marks since 
the precise meaning is that the {\em asymptotics} is perturbative.

We further report on our investigation of the numerical precision 
of perturbation theory for $P$ and then compare \eq{e:theequ}
to a first non-perturbative investigation for $\nq=2,\,\nl=0$,
which we expect to be a quite realistic model for real QCD.  
In this case, the lowest order effective theory is the Yang-Mills Theory,
as long as we look at the gluonic sector only, which we do here. 
Finally we argue through our numerical simulations that the  
effects of a charm quark, which are missed by simulating just QCD with $\nl$ 
quarks, are small in typical ratios of hadronic scales.

\section{The effective theory: \lowercase{dec}QCD }

The leading order low energy effective theory is 
${\rm QCD}_{\nl}$.  Next-to-leading order (NLO) correction terms 
in the local effective Lagrangian 
are gauge-, Euclidean- and chiral-invariant local fields.
These invariances allow only for fields, $\Phi_i(x)$, of at 
least dimension six.\footnote{To be precise, we here assume that 
either $\nl=0$ where light quark fields are absent, or $\nl\geq2$ 
where there is a non-anomalous chiral symmetry in the light sector. 
The statement also holds for theories with light quark masses if we define
light-quark mass-factors to be included in $\Phi_i(x)$. We thank 
Martin L{\"u}scher for a clarification on the use of chiral symmetry
in this context.} The Lagrangian may then be written as 
\bes
\lag{\rm dec}= \lag{QCD_\mathit{\nl}}
+ \frac1{M^{2}} \sum_i \omega_i \Phi_i +\rmO(M^{-4})\,,
\ees 
with dimensionless couplings $\omega_i$ which depend logarithmically on the 
mass $M$. 

At the lowest order in $1/M$, a single coupling\footnote{
Again we refer to the theoretical situation where the first $\nl$ 
flavors are mass-less. In general, also the  
light quark masses have to be matched.}, $\gbareff$,
is adjusted such that the low energy physics 
of ${\rm QCD}_{\nl}$ and ${\rm QCD}_{\nq}$ match for energies 
$E\ll M$. It then suffices to require one physical low-energy 
observable to match, e.g.\ a physical coupling. 
Discussing the issue in perturbation theory \cite{thresh:BeWe}, 
Bernreuther and Wetzel chose 
the MOM-coupling as a physical coupling and worked
out the matching of the $\msbar$ coupling. 
Meanwhile, the matching of the latter is 
known to high perturbative order. 
We use this information below. 

For now, we remain with the lowest order theory, i.e.\ all terms
$\rmO(E^2/M^2)$ are neglected and the Lagrangian is $\lag{\rm dec}=\lag{QCD_\mathit{\nl}}$.
We just make use of the fact that there 
is a single coupling, the gauge coupling $\gbareff$. Specifying a 
renormalisation scheme, its  $\beta$-function is fixed and the coupling 
is  a unique function $\gbareff = \gbarnl(\mu/\Laml)$, where $\mu$ is the renormalisation
scale. Therefore the matching condition between $\gbareff$ and $\gbarnf$ is equivalent
to a relation between the $\Lambda$-parameters. Considering only RGIs,
the only additional parameter is the quark mass $M$
of the fundamental theory. Therefore, we have to set 
\bes
   \Laml =  \Lameff(M,\Lamq)
\ees
in order to match the two theories. For dimensional reasons
the unknown function $\Lameff$ can be written as
\bes
    \label{e:lamrat}
   \Lameff(M,\Lamq) =  P_{\tl,\tq}(M/\Lamq) \, \Lamq \,.
\ees
In general the $\Lambda$-parameter of an asymptotically
free theory is a free, dimensionful, constant, which is to be fixed
from outside, usually by matching the theory to experiment. In the present case, experiment for
${\rm QCD}_{\nl}$ is replaced by ${\rm QCD}_{\nq}$ where the overall
energy scale $\Lamq$ remains free as before. 

The factorisation \eq{e:theequ} is a simple consequence of \eq{e:lamrat}: 
consider low energy 
scales of the theory, in particular hadron masses $\mhad$. 
After matching (and neglecting terms of order $\Lamq^2/M^2$)
they are equal in the fundamental and in the effective theory, 
$\mhad_\tl = \mhad_\tq$. We note further, that in ${\rm QCD}_{\nl}$
there are no mass parameters,
the only scale is $\Laml$ and hence
hadron masses are $\mhad_\tl = \rho^\mathrm{had} \Laml$ with pure 
numbers $\rho^\mathrm{had}$. 
Thus $\mhad_\tl/\Laml$ is independent of $M$.
In the fundamental theory 
$\mhad_\tq(M)/\Lamq$ does  of course depend on $M$, but 
$\Lamq$ is by definition independent of $M$. Together these facts 
entail the relation \eq{e:theequ} with
\bes
   Q^\mathrm{had}_{\tl,\tq} =  {\mhad_\tl/ \Laml \over 
                   \mhad_\tq(0)/\Lamq }  \,
\ees
defined entirely through the two mass-less theories.

Even though the physics of the two theories is matched at energy scales far below the mass, 
the perturbative  matching of the couplings is in fact 
best done with a renormalisation scale $\mu$ of the order of the mass 
\cite{Weinberg:1980wa,thresh:BeWe}. Higher order perturbative
corrections then vanish asymptotically as $M\to\infty$ and
the matching of the couplings is indeed 
perturbative. This entails that $P_{\tl,\tq}$ can be computed 
in perturbation theory when the mass is large. 

The  Bernreuther-Wetzel
relation between the $\msbar$ couplings
$\gbareff=\gbarnl(\mstar/\Laml)$ and $\gbarnf\equiv\gbarnf(\mstar/\Lamq)$
is meanwhile known to four loops \cite{Grozin:2011nk,Chetyrkin:2005ia},
\bes
    \label{e:matchgbar}
     \gbareff^2=\gbarnf^2 \times
   \left[ 1+c_2\gbarnf^4+c_3\gbarnf^6+\ldots \right]\,, 
\ees
where
$
   c_2 = (\nq-\nl) \, {11\over 72}\,(4\pi^2)^{-2}\,, 
$ 
and    
$
   c_3 =  
     \left[{\frac {572437}{62208}}
    -{\frac {84185}{13824}}\,\zeta_3 
    -{\frac {2633}{15552}}\,\nl\right]\,(4\pi^2)^{-3}
$ for \mbox{$\nq-\nl=2$}, and
$
   c_3 =  
     \left[{\frac {564731}{124416}}
    -{\frac {82043}{27648}}\,\zeta_3 
    -{\frac {2633}{31104}}\,\nl\right]\,(4\pi^2)^{-3}
$ for $\nq-\nl=1$. In this relation, the $c_1\gbarnf^2$ term in the brackets
is missing since $c_1$ vanishes for our choice of renormalisation
scale, $\mu=\mstar$, where $\mstar$ satisfies $\mbar_\msbar(\mstar/\Lamq)=\mstar$
with $\mbar_\msbar$ the quark mass in the $\msbar$ scheme.

From now on we suppress indices $\tl,\tq$ on $\Lambda$ and $\gbar$, since 
the effective theory only appears implicitly through the previously defined 
quantities $Q,P$.
We define a renormalisation group invariant mass scaling function
by the logarithmic derivative ($P'(x)=\frac{\rmd}{ \rmd x}P(x)$)
\bes\label{e:etaM}
  \etargi(M) \equiv  {M\over P }\! \left.{\partial P \over \partial M}\right|_\Lambda\!
  = {M\over \Lambda}{P'\over P}\!
  \simas{M\to\infty}\, \eta_0 + \etargi_1 \gbar^2  + \ldots\;\phantom{,}
\ees
with respect to the RGI mass $M$. Just like $M$ itself, $\etargi(M)$
is independent of the scheme. Residual dependences only result when it is 
evaluated approximately, e.g.\ at a finite order of perturbation theory.
We worked out its perturbative expansion \cite{decouppaper}, using \eq{e:matchgbar} and
the known expansions of the QCD $\beta$-function and the mass anomalous dimension in the
$\msbar$ scheme up to 4-loop \cite{MS:4loop1,MS:4loop2,MS:4loop3,Czakon:2004bu}.
Here we only report 
\bes
     \eta_0 &=& 1 -{\bzero \over \bzerop} > 0 \,,\\
     \etargi_1 &=& -{\bzero \over \bzerop}\left({\bone \over \bzero} -{\bonep \over \bzerop}\right)
                   - {\eta_0 \over 2\pi^2}\,,
\ees
with $b_0(n)=(11-2n/3)/(4\pi)^2$, $b_1(n)=(102-38n/3)/(4\pi)^4$ and refer the reader
to \cite{decouppaper} for the 
general expressions and details of the perturbation theory. 
Integrating \eq{e:etaM} gives an asymptotic
expression ($\logML=\log(M/\Lambda)$)
\bes\label{e:P}
  P = \frac{1}{k} \exp(\eta_0\logML)\; \logML^{\etargi_1/2\bzero} \times
        \left(1 + \rmO\big(\frac{\log\logML}{\logML}\big) \right),
\ees
where the constant $k$ is fixed by our conventions for the $\Lambda$ parameter and the RGI mass $M$
\cite{mbar:pap1} to
$
     \log k = {\bone \over 2 (\bzero)^2}\log 2 - {\bonep \over 2 (\bzerop)^2}
                       \log(2\bzero/\bzerop)
$.
It turns out that in the $\msbar$ scheme 
the higher order corrections to $\etargi$ as well as the 
function $P$ are very small as far as they are known, namely up to an impressive 4-loop level,
$\etargi_3\,g^6$. We discuss an example below.

We now turn to a non-perturbative investigation of the question how
well the mass-dependence at intermediate masses $M$ matches onto the 
asymptotic perturbative prediction. For this purpose we simulate a model,
namely QCD with two heavy, mass-degenerate quarks. The effective theory,
\qcdm, then is the Yang-Mills
theory up to $1/M^2$ corrections ($\nq=2$, $\nl=0$).

In Monte Carlo simulations of QCD with $\nf=2$ mass-degenerate
O($a$) improved Wilson fermions \cite{impr:csw_nf2} we
compute hadronic scales, e.g. $r_0(M)/a$,
at three values of the lattice spacing $a=0.066\,\fm$, $0.049\,\fm$ and $0.034\,\fm$.
The RGI mass $M$ is obtained along the lines of
\cite{alpha:lambdanf2}.
For details about the numerical computations, performed with MP-HMC \cite{lat10:marina},
openQCD \cite{algo:openQCD} and the package \url{https://github.com/to-ko/mesons}, 
and the methods applied we refer to \cite{decouppaper,lat14:Francesco}.

\begin{figure}[t]
   \includegraphics*[angle=0,width=\linewidth]{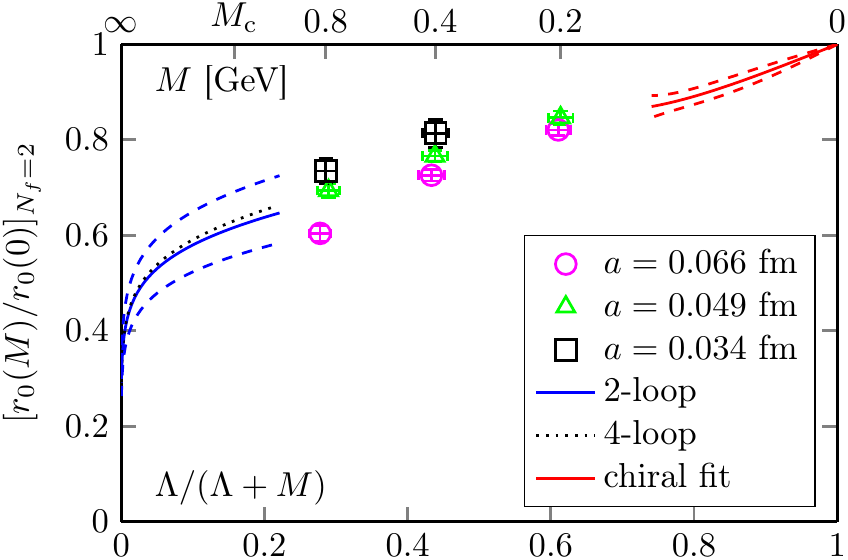}
  \caption{\label{f:bigplot}The mass-dependence of $r_0(M)/r_0(0)$ in the $\nf=2$ theory.
Monte Carlo data (symbols) are compared with the perturbative predictions for
$1/(QP)$ at large $M$. A fit to data close to the chiral limit is also shown.}
\end{figure}
For the hadronic scale $r_0$ \cite{pot:r0}, \eq{e:theequ} takes the form
$
  {r_0(0) / r_0(M)} =  Q
           \times {P(M/\Lambda) } + \rmO((\Lambda/M)^2)
$
with $Q={\left[\Lambda r_0(0)\right]_{\nf=2} / \left[\Lambda r_0\right]_{\nf=0} }$.
The ratios $r_0(M)/r_0(0)$ for $\nf=2$ are plotted in 
\fig{f:bigplot} as a function of $\Lambda/(\Lambda+M)$.
The value $r_0(0)/a$ in the chiral limit is taken from \cite{alpha:lambdanf2}
for $a=0.066\,\fm$ and $0.049\,\fm$,
and we estimate it to $13.06(42)$ at $0.034\,\fm$.

The red curve in \fig{f:bigplot} shows the mass-dependence close to 
the chiral limit as
fitted in \cite{alpha:lambdanf2} with the dashed red lines representing
the error of the fit.
At large $M/\Lambda$ the blue curve in \fig{f:bigplot} is drawn using the 2-loop perturbative
formula for $P$ in \eq{e:P} and
$Q = 0.789(52)/0.602(48) = 1.30(14)$ known from previous work \cite{mbar:pap1,alpha:lambdanf2}.
The dashed blue lines represent the uncertainty of $Q$.
The dotted black curve is drawn using the 4-loop value of $P$ and shows that
higher perturbative orders
are very small. They are negligible in comparison to the uncertainty of $Q$.
As our present non-perturbative results, we take the values 
at the smallest lattice spacing ($a=0.034\,\fm$).  
For $M/\Lambda=2.50$ or $M\approx0.8\,\GeV$, a rather modest value of the mass, these are consistent with 
the (upper error bar of the) factorisation curve. Thus within our 
precision, the perturbative prediction is verified. 

By discretizing the derivative in \eq{e:etaM} as $\etargi\approx 
\log(r_0(M_2)/r_0(M_1)) / \log(M_2/M_1)$ we obtain from our
simulations numerical estimates of $\etargi$. Their values are between
0.12 and 0.17 and are very close to perturbation theory, $\eta_0\approx 0.12$. 
A more precise statement 
needs a careful continuum limit, both for $\etargi$ and in
\fig{f:bigplot}. The lattice
community should address this issue in the near future.
%
\begin{figure}[t]
   \includegraphics[angle=0,width=\linewidth]{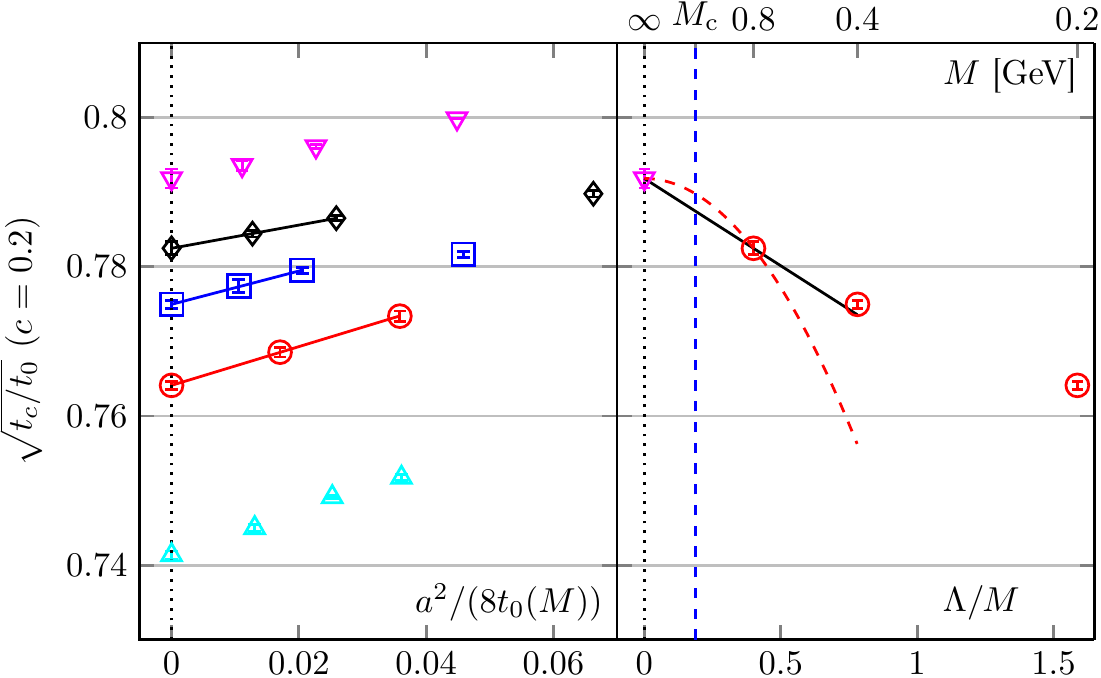}
  \caption{\label{f:tcot0}Left: The continuum extrapolation of the ratio $\sqrt{t_c/t_0}$
($c=0.2$) at mass values (from top to bottom) $M/\Lambda = \infty$, 2.5, 1.28, 0.63, 0.
Right: Its mass-dependence including a linear and quadratic interpolation in $\Lambda/M$
between the largest mass and $\nf=0$ ($M/\Lambda=\infty$).
}
\end{figure}
%

\section{Power corrections $\rmO(\Lambda^2/M^2)$}

So far we have discussed a comparison of the full theory to the prediction
of the factorisation formula resulting from the lowest order effective theory. 
When we take ratios of different hadron masses or different hadronic scales,
the function $P(M/\Lambda)$ drops out and we have access to the 
$\rmO(\Lambda^2/M^2)$ power corrections without any perturbative uncertainties.
We consider ratios
\bes
 \label{e:ratios}
R = \sqrt{t_0}/w_0 \,,\; r_1/r_0\,,\; r_0/\sqrt{t_0}\,,\; \sqrt{t_c/t_0}
\ees 
where the scale $t_c$ is defined through the 
smoothed action density \cite{flow:ML} $E(t)$ via $t_c^2\langle E(t_c)\rangle = c$
with $c=0.2$. It is a shorter distance cousin of $t_0$ \cite{flow:ML}.

We target the mass values $M/\Lambda = 0.63$, 1.28, 2.50
which correspond approximately to 0.2, 0.4, 0.8 $\GeV$. For comparison
the RGI charm mass $\Mc\approx 1.6\, \GeV$ \cite{PDG-2014}. We
correct the ratios $R$ for small differences
between the targeted and the simulated values of the masses.
In the corrections we neglect the error on $M/\Lambda$ since it mainly 
comes from $\Lambda$ and is therefore common to all points.

Our continuum extrapolations are performed by global fits, 
\bes
 \label{e:globalfit}
R_\mathrm{Lat} = R(M) + s \frac{a^2}{8t_0}\left(1 + k_1 \frac{M}{\Lambda} + k_2 \frac{M^2}{\Lambda^2} \right) \,,
\ees
to all the data.
Where it is known, we fix the slope $s$ (which describes the
mass independent cut-off effects) from its value determined at $M=0$, cf.
\cite{Sommer:2014mea}.
As a representative case,
we show in \fig{f:tcot0} (left) the global fit for $R=\sqrt{t_c/t_0}$. The slope
$s=0.295$ has been determined from a continuum extrapolation of $\sqrt{t_c/t_0}$ 
in the chiral limit (cyan upward-facing triangles). Our fits yield $k_2$ compatible
with zero. We drop it for our preferred continuum extrapolation, which 
then gives $k_1=-0.19(6)$ and an excellent quality of the fit.
The continuum limit values are very precise and allow to determine the size
of the mass effects in the ratio $R$.
For comparison, the magenta downward-facing triangles in \fig{f:tcot0} are the
results for $\nf=0$, which according to \eq{e:theequ} is recovered in the limit $M/\Lambda\to\infty$.

In \fig{f:tcot0} (right) we plot the values $R(M)$
(red circles) together with $R(\infty)$ in the $\nf=0$ Yang-Mills 
theory (magenta downward-facing triangle).
While the effective theory expectation is a roughly quadratic behavior in 
$\Lambda/M$, the full theory results are approximately linear in that 
variable. The natural explanation -- since we do not have any doubt about the validity of the effective theory description --
is that the masses of our simulations are not yet large enough to be
described by NLO \qcdm\ (Yang-Mills plus $1/M^2$ corrections). 
Taking the largest mass and
the $\nf=0$ value 
we can obtain by simple linear interpolation in $1/M$ (black line) and
$1/M^2$ (red dashed line)
two estimates of the mass effects at the charm mass marked by the
blue vertical dashed line.
\begin{table*}[t]
\begin{ruledtabular}
\begin{tabular}{ccdddddd}
& $M/\Lambda\to$ & \multicolumn{2}{c}{$\Mc/\Lambda$} & 2.50 & 1.28 & 0.63  & 0\\
$R$ && \multicolumn{1}{c}{$1/M$-scaled} & \multicolumn{1}{c}{$1/M^2$-scaled} & & & & \\\hline
$\sqrt{t_0}/w_0$ && 0.34(5)\% & 0.16(2)\% & 0.72(11)\% & 1.26(12)\% & 2.62(14)\% & 5.4\%\\
$\sqrt{t_c/t_0}$ && 0.28(3)\%  & 0.13(1)\% & 0.59(6)\% & 1.06(3)\% &  1.74(3)\% & 3.2\% \\
$r_1/r_0$        && 0.45(13)\% & 0.21(6)\% &  1.0(3)\%  &  1.8(5)\%  &  2.6(6)\% & \approx 4.0\%\\
$r_0/\sqrt{t_0}$ && 0.05(28)\% & 0.02(12)\% &  0.1(6)\%  &  0.7(5)\% & 1.7(5)\% & 3.0\% \\
\end{tabular}
\end{ruledtabular}
\caption{\label{t:releff}Relative effects \eq{e:releff} for the ratios in \eq{e:ratios}.
At $\Mc$ we quote the results from interpolations in $1/M$ 
and $1/M^2$, see \fig{f:tcot0}.}
\end{table*}

The dynamical fermion effects of these heavy quarks are very small 
and it is hence expected that they are
strongly dominated by the contribution of a
single fermion-loop (but non-perturbative in $\gbar$ and after renormalisation). 
As a result one expects a rather linear dependence on $\nf$. Since the relevant effect
for physics is the contribution of a single heavy quark,
we rescale the relative mass effect as $(\Nf=2)$
\bes
 \label{e:releff} 
    {1 \over \Nf} {R(M)  - R(\infty) \over R(\infty)}\,.
\ees
These numbers are listed in \tab{t:releff} for the ratios
in \eq{e:ratios}.

\section{Conclusions}

In conclusion, we pointed out the factorisation formula \eq{e:theequ} for the dominating
dependence of low energy dimensionful
quantities such as hadron masses on the mass of a heavy (dynamical) quark.
In perturbation theory, the power law $P \sim (M/\Lambda)^{\eta_0}$ is a 
very good approximation and we find that the non-perturbative
dependence is also rather close to that law for quark masses around 
$\frac12 M_\mathrm{c} \ldots \frac14 M_\mathrm{c}$. The knowledge 
of this mass-dependence is expected 
to provide valuable  information for tuning heavy quark masses to 
the correct point in future lattice QCD computations. 
We emphasise that our results are entirely sufficient to get the qualitative picture.
At the quantitative level, they are limited to
an accuracy of around 10\%, both because of the limited precision
in the mass-less theory and because we have not yet taken a true 
continuum limit for the finite mass points in \fig{f:bigplot}.
At least the latter should be improved soon. In principle 
one also has to worry about power corrections to the factorisation formula,
but \tab{t:releff}
shows that these are irrelevant at the present level of precision.

The dominating effect in \fig{f:bigplot} originates from the 
mass-dependence of the gauge coupling in the effective theory. 
It therefore disappears
in dimensionless ratios of low energy scales at fixed mass $M$ and only leaves 
residual power law effects. The effective theory analysis predicts
those to be of the form $M^{-2}$ for large $M$. Our investigation
of these power corrections has been restricted to 
$M \leq \frac12 M_\mathrm{c}$.
Larger masses require smaller lattice spacings, larger lattices 
and (due to critical slowing down) larger statistics. However,
in the accessible region we have precise results. Phenomenologically 
they are described by an approximate $M^{-1}$ law. 
We therefore interpolated between the largest simulated 
mass and the Yang-Mills theory to the charm mass 
as  $M^{-n}$ with both $n=1$ and  $n=2$. 
It seems safe to assume that the 
true results will be in between. In any case, the thus interpolated 
effects are very small, between 1 and 6 permille (\tab{t:releff}).
This provides a message for today's dynamical fermion simulations.
Dynamical charm effects are relevant only when one has very good precision, 
a very small lattice spacing and/or physical observables sensitive 
to higher energy scales.

\begin{acknowledgments}
We thank Hubert Simma and Ulli Wolff for their valuable comments on
the manuscript and Martin L\"uscher for a very useful discussion on the effective theory.
We thank Andreas Athenodorou and Marina Marinkovic for their contributions
in an early stage of this work.
Simulations have been performed on the supercomputers
Konrad and Gottfried at HLRN,
Cheops at the University of Cologne (financed by the Deutsche Forschungsgemeinschaf),
Stromboli at the University of Wuppertal
and PAX at DESY, Zeuthen.
We thank these institutions for support.
This work is supported by the Deutsche Forschungsgemeinschaft
in the SFB/TR~09 and the SFB/TR~55. 
\end{acknowledgments}

\bibliographystyle{apsrev4-1} 
\bibliography{latticen,qcd,latticen_fk,qcd_fk,partphys}

\end{document}